\documentstyle[12pt]{article}

\begin{document}

\hfill{UTTG-12-02}

\vspace{36pt}

\begin{center}
{\large {\bf { Adiabatic Modes in Cosmology}}}

\vspace{36pt}
Steven Weinberg\footnote{Electronic address:
weinberg@physics.utexas.edu}\\
{\em Theory Group, Department of Physics, University of
Texas\\
Austin, TX, 78712}

\vspace{30pt}

\noindent
{\bf Abstract}
\end{center}

\noindent
We show that the field equations for cosmological perturbations in 
Newtonian  gauge always have an  
adiabatic solution, for which a  quantity ${\cal R}$ is  non-zero and
constant in all eras in the limit of large wavelength, so that it can
be used to connect observed cosmological fluctuations in this mode 
with those at very early times. There is also a second adiabatic mode, 
for which ${\cal R}$ vanishes for large wavelength, and in general there 
may be non-adiabatic modes as well.  These conclusions apply in all eras and
whatever the constituents of the universe, under only a mild technical
assumption about the wavelength dependence of the field equations 
for large wave length. In the absence of anisotropic inertia, 
the perturbations in the adiabatic modes are given for large wavelength
by universal formulas in terms of the Robertson--Walker scale factor.
We discuss  an apparent discrepancy between 
these  results and what appears to be a conservation law in all modes
 found for large wavelength in synchronous gauge: it turns out that,
although equivalent, synchronous and Newtonian gauges suggest 
inequivalent assumptions about the behavior of the perturbations for 
large wavelength.

 \vfill

\pagebreak
\setcounter{page}{1}

\begin{center}
{\bf I. INTRODUCTION }
\end{center}

If observations are to be used to tell us something about inflation, 
then we need some way of connecting the properties of the cosmological 
fluctuations produced during inflation to the properties of 
fluctuations much closer to the present.  Inconveniently, in 
inflationary cosmologies the  era of inflation was followed by a 
period 
when the energy in scalar fields was converted into matter and 
radiation, and about this process we know essentially nothing.  
Subsequently there may have been other periods about which we are 
equally ignorant, such as the often-hypothesized era with temperatures 
between $10^{13}$ GeV and $10^{11}$ GeV, when supersymmetry may have 
become broken by unknown strong forces.  These mysterious eras 
occurred 
when fluctuations of cosmological interest were far outside the 
horizon, but this does not 
rule out some effect on the strength or even the wave-length 
dependence of these 
fluctuations.\footnote{By a mode being ``beyond the horizon'' we only 
mean 
that the physical  wave number is much less than the expansion rate.  
This does not necessarily have anything to do with causality; indeed, 
the point of inflation is to make the true particle horizon 
radius much larger than the inverse expansion rate.}  
Therefore, in relating the cosmological 
fluctuations produced during inflation with those observed in the 
cosmic microwave background or in 
large-scale cosmic structures, 
it is essential to employ some sort of 
conservation law that is valid at large wavelengths whatever the 
details of cosmic evolution. 

In much  work on fluctuations in cosmology, the 
conserved 
quantity is taken to be a quantity 
${\cal R}$ related to the spatial curvature on co-moving spatial 
surfaces[1],   given in Newtonian gauge by\footnote{Here $H=\dot{a}/a$ 
is the expansion rate, with  dots denoting  ordinary time derivatives.
In Newtonian gauge the perturbations to the
gravitational field are taken to be $\delta g_{00}=-2\Phi$ and 
$\delta g_{ij}=-2a^2\Psi\delta_{ij}$.  Also $\delta\rho$, $\delta p$,
and $\delta u$ are the perturbations to the total energy density, 
pressure,
and velocity potential in Newtonian gauge, while we use a bar to
denote unperturbed quantities like the unperturbed energy density
$\bar{\rho}$ and pressure $\bar{p}$. For simplicity we assume 
a vanishing unperturbed spatial curvature.}
\begin{equation}
{\cal R}=-\Psi+H\delta u\;.
\end{equation} 
The rate of change of 
${\cal R}$ is given by a general 
formula[2]: 
\begin{equation} 
\dot{\cal 
R}=X+\left(\frac{q^2}
{a^2}\right)\left[\left(\frac{\ddot{H}+3H\dot{H}}{3\dot{H}^2}\right) 
\Psi-
\left(\frac{H }{\dot{H}}\right) 4\pi G\,\delta\sigma \right]\;.
\end{equation} 
Here $q$ is the co-moving wave number, $\delta\sigma$ is a measure of 
the
anisotropic stress,\footnote{The quantity $\delta\sigma$ is defined
by writing 
$ T_{ij}$ for scalar perturbations as 
$g_{ij}p+\partial_i\partial_j\delta\sigma$. 
In this formalism, $p$ includes effects of bulk viscosity, 
while $i{\bf q}\delta u$ is the velocity of energy transport, and so
includes effects of heat conduction.  For this formalism, see ref. 
[3].} and  
\begin{equation} 
X\equiv \frac{\dot{\bar{\rho}}\,\delta p-\dot{\bar{ p }}\,\delta 
\rho}{3(\bar{\rho}+\bar{p})^2}\;.
\end{equation} 
Thus ${\cal R}$ is conserved in the limit of small wave number 
if and only if $X=0$ in this limit.

The limit of small $q$ is of some interest in itself, but its 
importance
lies chiefly in the circumstance that those factors of $q$ that arise 
from
the field equations (rather than from the initial conditions) are 
always
accompanied by factors of $1/a(t)$, because it is only $q/a(t)$ that 
is 
independent of the units chosen for the co-moving spatial coordinates 
$x^i$.
It is usually a good guess that terms in the perturbations 
proportional to 
such factors of $q$ will be negligible if $q/a(t)H(t)\ll 1$. Hence, 
although 
here we will study the behavior of the perturbations
in the limit of small $q$, it is expected that this
provides insight to the behavior of perturbations as $a(t)$ increases.
In particular, we expect that if
$X\rightarrow 0$ for $q\rightarrow 0$, and if the coefficient of $q^2$ 
in the 
second term in Eq.~(2) remains finite in this limit, then at any given 
epoch 
$\dot{\cal R}$ will be small if $q/aH$ is sufficiently small.  In any 
mode for
which $\cal R$ is non-zero in this limit, the fractional
rate of change of ${\cal R}$ will then also be small in this limit.

Now, the quantity $X$ vanishes in all modes and for all $q$ when the 
perturbed pressure 
$\bar{p}+\delta p$ is a function only of the perturbed energy density 
$\bar{\rho}+\delta\rho$, as is the case in a universe dominated 
either by pure radiation
or by pure cold matter, but not when both radiation and cold matter 
are 
important,  and 
also not  during inflation  or in the curvaton 
model[4].  The quantity $X$ does vanish for all modes in the limit 
$q\rightarrow 0$ in the case
of inflation with a single scalar field, but this is not true with 
several 
scalar fields.
Section II of this paper will show that in 
general, 
whatever the contents 
of the universe,  with only a mild technical assumption about the 
dependence on wave number of the field equations for 
cosmological perturbations in 
Newtonian  gauge, these equations always have 
a physical solution for which 
$X\rightarrow 0$ and ${\cal R}$ approaches a non-zero constant in the 
limit  $q\rightarrow 0$, though there may also be 
other  modes for which $\cal 
R$
is not constant. 
In fact, there are
always at least two  solutions with $X=0$ and $\cal R$ constant in
the limit $q\rightarrow 0$: one solution for which ${\cal R}\neq 0$, 
and another with ${\cal R}=0$. These  solutions  
will be illustrated in Section III for the case of 
inflation  with any number of interacting scalar fields.
The existence of such  solutions is well 
known in special cases, but I do not know of a previous general 
proof of their existence.

Solutions of this sort are usually called {\em adiabatic}, even in 
contexts
where thermodynamics has no 
relevance.\footnote{Sometimes the non-adiabatic solutions of the 
field equations are called isocurvature perturbations.  This is a 
misnomer,
because even for $q=0$ it is only possible for ${\cal R}$ to have the
constant value zero if $X=0$.  Nevertheless, as we will see in Section 
IV,
in synchronous gauge there is a sense in which the solutions that do 
not 
correspond to the adiabatic solutions 
of Newtonian gauge can indeed be regarded as isocurvature modes.}  
As we will also see in Section II, in theories in which the energy-
momentum
tensor is the sum of a number of tensors $T_f^{\mu\nu}$
for  a set of fluids labeled $f$, we have the stronger result that 
for $q=0$ in the adiabatic modes, the perturbations in each fluid
satisfy
\begin{equation} 
\frac{\delta\rho_f}{\dot{\bar{\rho}}_f}=\frac{\delta 
p_f}{\dot{\bar{p}}_f}=
\frac{\delta\rho}{\dot{\bar{\rho}}}=\frac{\delta p}{\dot{\bar{p}}}\;.
\end{equation} 
In the special case where the unperturbed energy-momentum 
tensors are separately conserved, 
we also have $\dot{\bar{\rho}}_f=-3H(\bar{\rho}_f+\bar{p}_f)$, 
in which case
Eq.~(4) implies that 
 the ratios $\delta\rho_f/(\bar{\rho}_f+\bar{p}_f)$ are 
equal:
\begin{equation} 
\frac{\delta\rho_f}{\bar{\rho}_f+\bar{p}_f}=
\frac{\delta\rho}{\bar{\rho}+\bar{p}}\;,
\end{equation} 
which is often taken to be what is meant by an adiabatic
perturbation.

Things appear very different in synchronous gauge. As 
shown in Section IV, when we take the limit of vanishing wave number
in the field equations of synchronous gauge, we find that {\it these 
equations respect the conservation of  a quantity $A$ in all modes 
whatever
the contents of the universe}, provided only that none of the 
perturbations in synchronous
gauge blow up in the limit $q\rightarrow 0$. 
At first sight, this presents an apparent paradox.  All gauges are 
 equivalent, so how can there be a quantity that is conserved 
for zero wave number in all modes during all eras in synchronous  
gauge 
and no equally universal conservation law for zero wave number in 
Newtonian gauge?  We will find the answer to be that when we 
speak of the limit $q=0$ we mean different things in different gauges.
Though mathematically equivalent, synchronous gauge and Newtonian 
gauge
suggest different hypotheses about how perturbations behave in this 
limit,
leading to different conditions for the validity of the conservation
law for $q=0$.

In some work on cosmological fluctuations, a quantity $\zeta$, related 
to the spatial curvature on spacelike surfaces of constant energy 
density, is used in place of ${\cal R}$.  It is defined in Newtonian 
gauge by[5]
\begin{equation}
\zeta\equiv -\Psi-H\delta\rho/\dot{\bar{\rho}}=-
\Psi+\delta\rho/3(\bar{\rho}+\bar{p})\;.
\end{equation} 
By taking a suitable linear combination of the time-time and time-
space 
components of the Einstein field equation and the part of the space-
space
component proportional to $\delta_{ij}$, one can derive a general 
constraint[6],
\begin{equation} 
a^3\delta\rho -3Ha^3\left(\bar{\rho}+\bar{p}\right)\delta u 
+\left(\frac{q^2a}{4\pi G }\right)\Psi=0\;,
\end{equation} 
so that
\begin{equation}
{\cal R}-\zeta=\left(\frac{q^2}{3a^2\dot{H}}\right)\Psi
\end{equation} 
and therefore in all modes $\zeta \rightarrow {\cal R} $ in the limit 
$q\rightarrow 0$.

\begin{center}
{\bf II. ADIABATIC MODES IN NEWTONIAN GAUGE}\nopagebreak 
\end{center}

We consider a general cosmological model, based on the Einstein field 
equations,
supplemented with whatever other equations are needed to give the 
energy 
density,
pressure, and velocity potential perturbations in terms of independent 
dynamical
variables, and the field equations satisfied by those variables.  We 
will 
demonstrate the general existence of a pair of adiabatic 
solutions of these field
equations in Newtonian gauge: one with ${\cal R}\neq 0$ and constant 
in the limit $q\rightarrow 0$, 
and the other with ${\cal R}=0$ in this limit.  In order to draw these
conclusions  without specifying the formulas for $\delta\rho$, $\delta 
p$,
and $\delta u$, we use a trick, based on the fact that although there 
is no
remaining gauge symmetry in Newtonian gauge for $q\neq 0$, for $q=0$ 
there is
 a remnant gauge symmetry of the field
equations in Newtonian gauge, which makes it easy to find exact 
general
solutions of the field equations for $q=0$.  Not all of these
solutions are physical.  To be physical, such
a solution must be the limit as $q\rightarrow 0$ of a solution of the 
field equations for $q\neq 0$ in at least a neighborhood of $q=0$.
For this to be the case, it is necessary for the $q=0$ solution to 
satisfy
certain conditions imposed by the Einstein field equations, which 
limit the 
physical solutions to a linear combination of just 
two independent adiabatic 
modes.  
We will then make a mild technical
assumption about the dependence of the 
field equations on $q$, which guarantees
that these two solutions are the limits as $q\rightarrow 0$ of 
physical
solutions 
for $q\neq 0$.

Whatever the constituents of the universe, for $q=0$ the field 
equations in Newtonian
gauge for scalar (i.e., compressional) modes will always be invariant 
under the gauge transformation induced by a redefinition
of the time coordinate
\begin{equation} 
t\rightarrow t+\epsilon(t)\;,
\end{equation}
and a re-scaling of the space coordinate
\begin{equation} 
x^i\rightarrow x^i(1-\lambda)\;,
\end{equation}  
with $\epsilon(t)$ an arbitrary infinitesimal function of time
and $\lambda$ an arbitrary infinitesimal constant.  For scalar modes, 
$\delta g_{i0}$ 
is proportional to $q_i$, and the part of $\delta g_{ij}$ not 
proportional
to $\delta_{ij}$ is proportional to $q_iq_j$, so both automatically 
vanish 
for $q\rightarrow 0$, and we therefore do not need to impose any 
conditions on
$\epsilon(t)$ and $\lambda$ to remain in Newtonian gauge for $q=0$.  
Eqs.~(9) 
and (10) provide the most general spacetime transformations of purely
scalar perturbations that preserve the condition $q=0$.

Under this gauge transformation, the $q=0$ perturbations undergo the 
transformation
\begin{equation} 
\Psi\rightarrow\Psi+H\epsilon-\lambda\;,~~~~~\Phi\rightarrow\Phi-
\dot{\epsilon}\;,
\end{equation} 
\begin{equation} 
\delta\rho\rightarrow \delta\rho-\dot{\bar{\rho}}\epsilon\;,~~~~~~
\delta p\rightarrow \delta p-\dot{\bar{p}}\epsilon \;.
\end{equation}
Likewise, in the case of separate fluids with energy density and 
pressure $\rho_f$ and $p_f$
or separate scalar fields $\varphi_f$, 
\begin{equation} 
\delta\rho_f\rightarrow \delta\rho_f-
\dot{\bar{\rho}}_f\epsilon\;,~~~~~~
\delta p_f\rightarrow \delta p_f-\dot{\bar{p}}_f\epsilon \;,
\end{equation}
or
\begin{equation} 
\delta \varphi_f\rightarrow \delta \varphi_f-
\dot{\bar{\varphi}}_f\epsilon \;.
\end{equation} 
It follows that there is always a solution of the Newtonian 
gauge field equations for $q=0$, 
in which:
\begin{equation} 
\Psi=H\epsilon-\lambda\;,~~~~~~~~\Phi=-\dot{\epsilon}\;,
\end{equation} 
\begin{equation} 
\delta\rho=-\dot{\bar{\rho}}\epsilon\;,~~~~~~
\delta p=-\dot{\bar{p}}\epsilon \;,
\end{equation}
and for several fluids
\begin{equation} 
\delta\rho_f=-\dot{\bar{\rho}}_f\epsilon\;,~~~~~~
\delta p_f=-\dot{\bar{p}}_f\epsilon \;,
\end{equation}
or several scalar fields
\begin{equation} 
\delta \varphi_f=-\dot{\bar{\varphi}}_f\epsilon \;,
\end{equation} 
where $\epsilon(t)$ is an arbitrary function of time and $\lambda$ is 
an
arbitrary constant.  (It is not 
necessary for this that the 
energy-momentum tensors of the individual fluids or
scalar fields be separately conserved; all we need is that they are
tensors.)

Of course, for general $\epsilon(t)$ and $\lambda$ this is just a 
gauge
mode.  For it to have any physical significance, it must satisfy 
certain
conditions that allow it to be extended to the case of non-zero wave 
number.
In particular, the part of the space-space 
component of the Einstein field equations that is not proportional to 
$\delta_{ij}$ takes the form (with $\delta\sigma$ the anisotropic 
stress):
\begin{equation}
q_iq_j\Big(\Phi-\Psi)=-8\pi G q_iq_j\delta\sigma\;,
\end{equation} 
so this equation disappears for $q=0$.  In order for the solution 
(15)--(18) 
of the field equations to be extended to $q\neq 0$,
we must have 
\begin{equation} 
\Phi=\Psi-8\pi G\delta\sigma\;,
\end{equation}  and therefore $\epsilon(t)$ and 
$\lambda$ must satisfy the condition
\begin{equation}
\dot{\epsilon}+H\epsilon=\lambda-8\pi G\delta\sigma\;,
\end{equation} 
which can always be satisfied by a suitable choice of $\epsilon(t)$.  

There is another equation that 
disappears for $q=0$: for scalar modes
the space-time component of the 
Einstein field equations reads
\begin{equation} 
q_i\Big(-2\dot{\Psi}-2H\Phi\Big)=8\pi G 
\Big(\bar{\rho}+\bar{p}\Big)\,q_i\delta u =
-2\dot{H}q_i\delta u
\end{equation} 
where $\delta u$ is the velocity potential in Newtonian gauge, which 
does not 
appear in the equations for $q=0$.  Hence in order for the solution we 
have found to
be extended to $q\neq 0$, this solution must also have a velocity 
potential given by
\begin{equation} 
\dot{H}\delta u =\dot{\Psi}+H\Phi \;,
\end{equation} 
or, using the result (15),
\begin{equation}
\delta u=\epsilon\;.
\end{equation} 
This agrees with what would be found from the gauge transformation 
induced by the coordinate 
transformation (9).  Likewise, with several fluids,
\begin{equation} 
\delta u_f=\epsilon
\end{equation} 
From Eqs.~(16) and (24) it follows that $
\delta\rho=3 H \left(\bar{\rho}+\bar{p}\right) \delta u$,
so the constraint equation (7) is automatically satisfied for
$q=0$ by
this solution.
By inserting Eqs.~(15), (16), and (24) in equations (1) and (6),
we now find that  for 
$q=0$, this solution has
\begin{equation} 
{\cal R}=\zeta=\lambda\;,
\end{equation} 
so ${\cal R}$ and $\zeta$ are indeed constant and equal for this 
solution in the limit
$q\rightarrow 0$, as was to be shown.  They are also non-zero, as long 
as we take $\lambda\neq 0$.

Now we have to ask what additional conditions are needed to ensure 
that
this solution is the limit as $q\rightarrow 0$ of a 
solution with $q\neq 0$.
In general, any closed set of linear homogeneous ordinary differential 
equations for a finite set of dependent variables  
can be put in the first-order 
form
\begin{equation}
\dot{y}_n(t)+\sum_m C_{nm}(t)y_m(t)=0\;.
\end{equation} 
(If some of the original set of field equations involve 
derivatives of higher 
than first order,
we can still write the equations in the form (27) by 
including  some derivatives of 
the field variables among the $y_m(t)$.)  This has the general 
solution
\begin{equation} 
y_n(t)=\sum_m \left[T\left\{\exp\left(-\int_{t_0}^t 
C(t')\,dt'\right)\right\}\right]_{nm}y_m(t_0)
\end{equation} 
with $t_0$ arbitrary, and with $T$ denoting a 
time-ordered product defined 
by a power series expansion of the
exponential, which for finite matrices is always convergent.  
The initial conditions may be subject to constraints like Eq.~(7),
which can be written
\begin{equation}
\sum_nc_ny_n(t_0)=0\;.
\end{equation} 
(Eq.~(7) is such a constraint, because the equations of 
energy and momentum 
conservation and
the gravitational field equation $\dot{\Psi}+H\Phi=\dot{H}\delta u$ 
imply that the left-hand
side of Eq.~(7) is time-independent.)

Our ``mild technical''
assumption is that, as long as the  Einstein equations (19) and (22) 
are written instead in the stronger form 
(20) and (23), the matrix elements $C_{nm}(t)$ and the constraint 
coefficients $c_n$ are
continuous functions of  $q$ in at least a 
neighborhood of $q=0$.  In this case 
the $y_n(t_0)$ that satisfy Eq.~(29) and the 
matrix in Eq.~(28) will also be
continuous in $q$, so that
any solution of Eq.~(27) for $q=0$ can be extended to a 
solution for $q\neq 0$
in a neighborhood of $q=0$ by using Eq.~(28) and (29) 
with the values of $C(t)$ and
$c$ for $q\neq 0$.  The next section shows the validity of 
this assumption in
one illustrative example: inflation with several scalar fields and an 
arbitrary potential.  Because $q$ generally enters the 
field equations and 
constraint equations in such a simple manner, we expect that 
this assumption 
will always be satisfied, and in any case it is easy to check 
for any specific 
model.
For $q\neq 0$, there is no remaining gauge freedom in 
Newtonian gauge (as there is for $q=0$), so the adiabatic 
solution found in this way will be the limit as $q\rightarrow 0$ of a 
physical solution, not a mere gauge mode.

We expect the anisotropic stress 
coefficient $\delta \sigma$ for a wide class of 
theories to be some linear 
combination of 
$\delta u$, $\delta\rho$, and $\delta p$, so that for the
solution (15)--(17), (24)--(25), $\delta\sigma$ may be written 
as $\delta\sigma= \epsilon\Sigma $, 
with $\Sigma(t)$ depending only on 
unperturbed quantities.  (For instance, a non-zero shear viscosity 
$\eta$  in an imperfect fluid 
gives  $\delta \sigma=-2\eta\,\delta u$, so here $\Sigma=-2\eta$.) 
In all such cases, Eq.~(21) has the general solution
\begin{equation} 
\epsilon(t)=\frac{\lambda}{\alpha(t)}\int^t \alpha(t')\,dt'
\end{equation} 
where \begin{equation} 
\alpha(t)\equiv a(t)\exp\left(8\pi 
G \int^t \Sigma(t')\,dt'\right)\;,
\end{equation} 
 and the lower limit on the integral in Eq.~(30) is arbitrary.  

There is also a second mode, corresponding to the possibility of 
shifting the lower limit
of the integral in Eq.~(30), for which $\epsilon(t)$ goes as
$\epsilon(t)\propto 1/\alpha(t)$.
Since shifting the lower bound on the integral in Eq.~(30) has no 
effect on
the value (26) of ${\cal R}$ and $\zeta$, this solution has
${\cal R}=\zeta =0$.

In the special case of vanishing anisotropic stress 
we have $\delta\sigma=0$, 
so here $\Phi=\Psi$, and $\alpha(t)$
is just the Robertson--Walker scale factor $a(t)$.  
The general solution of Eq.~(21) is then 
\begin{equation} 
\epsilon(t) =\frac{\lambda}{a(t)}\int^t a(t')\,dt'\;,
\end{equation} 
with an arbitrary lower limit.  This eventually increases in absolute 
value as $t$ for Robertson--Walker
scale factors that grow as any power of $t$, while in the other mode
$\epsilon(t)\propto 1/a(t)$ decreases with time.  Inserting the result 
(32) in Eqs.~(15)--(18) gives 
explicit results for the
perturbations in the gravitational field and various pressures and 
energy densities as 
functions of time.  

The results presented in this section can be 
interpreted in terms of what Liddle and Lyth in ref. [1] 
call a ``separate universe'' picture, which in one form 
or another has been used since the beginning of 
inflationary theory to deal with cosmological 
fluctuations in the case of a single scalar field.  For 
instance, Bardeen, Steinhardt, and Turner in ref. [5] 
gave what they called a `heuristic argument' that in this 
case any portion of the universe that is larger than the 
horizon $1/H$ but smaller than the physical perturbation 
wavelength $a/q$ would have to look like a separate 
unperturbed universe, with $\varphi+\delta\varphi$ 
following the unique evolutionary path of the scalar 
field, and with all of these separate universes therefore 
the same except for a variation in the time at which the 
scalar field satisfies some specific initial condition a 
few Hubble times after horizon exit.  As pointed out by 
Bardeen {\em et al.}, it follows then that 
$\delta\rho/\dot{\bar{\rho}}=\delta p/\dot{\bar{p}}$, and 
hence $X=0$, for $q/a\ll H$.  

There is a potential problem with this sort of argument, that
there are two fields involved, the inflaton and the gravitational
field, so that different separate universes might have different
ratios of these fields.  The argument of Bardeen {\em et al.} was 
formulated in a gauge in which it is unnecessary to 
consider fluctuations in the gravitational field, but it 
applies also to Newtonian gauge, because in this gauge 
the constraint (7) allows the gravitational potential to 
be expressed in terms of fluctuations in the scalar 
field.  But as we have seen in this section, in Newtonian 
gauge it is necessary not only to allow shifts in the 
time at which the scalar field reaches some given value 
after horizon crossing, but also to re-scale the 
co-moving coordinates used in each separate universe. In 
synchronous gauge there is no constraint like Eq.~(7) 
that allows us to express the gravitational field in 
terms of the scalar field, and so, as we will see in 
Section IV, the solutions even for inflation with a 
single scalar field do not satisfy $X=0$ in the limit 
$q\rightarrow 0$.

There is another potential problem, that  the 
equation of motion of the scalar field is a {\em second}-order 
differential equation, so that there are two 
independent solutions whose relative coefficients may 
vary from one separate universe to another. Bardeen {\em et al.}
and other authors avoid this problem by assuming 
that the scalar field experiences a period of ``slow roll'' inflation,
in which the differential equation satisfied by the scalar field is 
of first order, to a good approximation.   We have not had to make 
this 
assumption, for a reason already 
pointed out by Guth and Pi[7]:   the Wronskian of these two
solutions decays rapidly after horizon crossing, so that 
it is as if there were only one independent solution. (Guth and Pi 
considered the case of $H$ constant, but even with a 
time-dependent $H$ the
Wronskian still decays, though not precisely exponentially.)

In any case, it has always been clear that such ``separate universe'' 
arguments do not rule 
out non-adiabatic solutions in the case of several scalar 
fields, where ratios of the scalar fields may vary from one 
``separate universe'' to another.  The results of this 
section may be interpreted as the statement that in this 
and all other cases it is always possible to find an 
adiabatic solution of the field equations in Newtonian 
gauge in which the separate universes appear the same, 
except for a shift in the time coordinate and a 
re-scaling of the co-moving space coordinates.

\begin{center}
{\bf III. AN EXAMPLE: MULTIFIELD INFLATION}
\end{center}

For illustration, 
and to confirm the reasoning of the 
theorem of the previous section in a case where $X$ does 
not vanish for all modes, let us consider the case
of inflation with an arbitrary number of scalar fields 
$\varphi_f$, and with a general potential
$V$ that may include interactions among the various scalars.  
The energy-momentum 
tensor of the scalar fields has the perfect-fluid form,
so here $\sigma=0$, and $\Phi=\Psi$.
The field equations 
in Newtonian gauge are 
\begin{equation} 
\dot{\Psi}+H\Psi=4\pi G \sum_f 
\dot{\bar{\varphi}}_f\delta\varphi_f\;,
\end{equation}
\begin{equation} 
\delta\ddot{\varphi}_f+3H\delta\dot{\varphi}_f+\sum_{f'}
\frac{\partial^2 V(\bar{\varphi})}{\partial \bar{\varphi}_f
\partial \bar{\varphi}_{f'}} \,\delta
\varphi_{f'}+\left(\frac{q^2}{a^2}\right)\delta\varphi_f 
=-2\Psi\frac{\partial 
V(\bar{\varphi})}{\partial 
\bar{\varphi}_f}+4\dot{\Psi}\dot{\bar{\varphi}}_f\;,
\end{equation}
and the constraint (7) is here
\begin{equation} 
\left(\dot{H}+\frac{q^2}{a^2}\right)\Psi =4\pi G\sum_f \left(-
\dot{\bar{\varphi}}_f\delta\dot{\varphi}_f+
\ddot{\bar{\varphi}}_f\delta\varphi_f\right)\;.
\end{equation} 
We can write Eqs.~(33) and (34) in the form (27) by taking the 
$y_n$ to run over $\Psi$ and all $\phi_f$ and $\dot{\phi}_f$, in which
case the constraint (35) is of the form (29).  Here obviously 
$C_{nm}(t)$
and $c_n$ are continuous in $q$ in a neighborhood of $q=0$; in fact,
they are just linear functions of $q^2$.  Hence any solution
of Eqs.~(33)--(35) that we find for $q=0$ can be extended to a 
solution
for $q\neq 0$.

Let us try for a solution for $q=0$ in which all of the 
individual velocity potentials
$-\delta{\varphi}_f/\dot{\bar{\varphi}}_f$ are equal, so that
\begin{equation} 
\delta\varphi_f=-\dot{\bar{\varphi}}_f\delta u\;,
\end{equation} 
with the common value satisfying
\begin{equation} 
\delta \dot{u}=-\Psi\;.
\end{equation} 
Using the time-derivative of the unperturbed scalar field equation
\begin{equation} 
\ddot{\bar{\varphi}}_f+3H\dot{\bar{\varphi}}_f+
\frac{\partial V(\bar{\varphi})}{\partial \bar{\varphi}_f}=0\;,
\end{equation} 
we can put  Eq.~(34) for $q=0$ in the form
\begin{equation} 
\dot{H}\delta u+H\delta\dot{u}
+\delta\ddot{u}=0\;.
\end{equation} 
Also, the gravitational field equation (33) now reads $
\dot{\Psi}+H\Psi=\dot{H}\delta u$, which Eq.~(39) 
guarantees is  automatically
satisfied by the $\Psi$ given by Eq.~(37).
The general solution is
\begin{equation} 
\delta u=\frac{\lambda}{a}\int a\,dt\;,~~~~
\Psi=H\delta u-\lambda\;,
\end{equation} 
(with $\lambda$ an arbitrary constant), just as we found 
above in Eqs.~(15), (24), and (32). 
The perturbations to the  energy density 
and pressure of the $f$th field here are
\begin{equation}
\delta\rho_f=-\Psi\dot{\bar{\varphi}}^2
+\dot{\bar{\varphi}}\delta\dot{\varphi}
+\frac{\partial V}{\partial\bar{\varphi}_f}
\delta\varphi_f=-\Big(\Psi+\delta\dot{u}\Big)
\dot{\bar{\varphi}}_f^2-\dot{\bar{\rho}}_f\delta u=
-\dot{\bar{\rho}}_f\delta u\;,
\end{equation} 
and
\begin{equation}
\delta p_f=-\Psi\dot{\bar{\varphi}}^2
+\dot{\bar{\varphi}}\delta\dot{\varphi}
-\frac{\partial V}{\partial\bar{\varphi}_f}\delta\varphi_f
=-\Big(\Psi+\delta\dot{u}\Big)
\dot{\bar{\varphi}}_f^2-\dot{\bar{p}}_f\delta u
=-\dot{\bar{p}}_f\delta u\;,
\end{equation}
so this mode is adiabatic, in the sense that  $X\rightarrow 0$ 
for $q\rightarrow 0$.  
Inserting Eq.~(40) in Eq.~(1)
gives again ${\cal R}=\lambda$.

Once again, because of the freedom to shift the lower 
limit of the integral in
Eq.~(40), there are two adiabatic modes here, the second 
with $\delta u\propto 1/a$ 
and ${\cal R}=0$.  

For a single scalar field,  Eqs.~(33) and (34) are a third-order 
set of differential 
equations, and therefore have a third independent solution.  The third 
solution can also be found explicitly, and turns out to have
$\dot{\cal R}\propto 1/a^3\dot{H}$ for $q=0$, so this solution is not
adiabatic.   However, this third solution is eliminated by the 
constraint Eq.~(35), which as we have seen  in the previous section 
is automatically satisfied by any adiabatic solution, 
but is not satisfied by
the non-adiabatic solution of Eqs.~(33) and (34).  For $N$ scalar 
fields 
Eqs.~(33) and (34) have $2N+1$ independent solutions, of which two are 
adiabatic,
and one is eliminated by Eq.~(35), leaving $2N-2$ 
non-adiabatic solutions.

\begin{center}
{\bf IV. SYNCHRONOUS GAUGE}
\end{center}

We now turn to synchronous gauge.  With zero unperturbed spatial 
curvature, 
the perturbed 
metric has components
\begin{equation}
g_{ij}({\bf x},t)=a^2(t)\delta_{ij}+h_{ij}({\bf x},t)\;,~~~~g_{00}=-
1\;,~~~~g_{i0}=0\;,
\end{equation} 
with $h_{ij}$ a small perturbation.
We now assume for simplicity that the perturbed energy-momentum 
tensor takes the 
perfect-fluid form
\begin{equation} 
T_{\mu\nu}=p\,g_{\mu\nu}+(p+\rho)u_\mu u_\nu\;.
\end{equation} 
The unperturbed 
quantities $\bar{p}$ and $\bar{\rho}$ depend only on time, and the 
unperturbed velocity four-vector has components  $\bar{u}^0=1$, 
$\bar{u}^i=0$.  The normalization condition $u_\mu u^\mu=-1$ then 
requires that the velocity perturbation $\delta u^{(S)}_{\mu}$ is 
purely spatial.  (A
superscript $(S)$ is used to denote perturbed quantities in 
synchronous 
gauge.)
We consider only compressional modes, for which $\delta 
u^{(S)}_i=\partial 
\delta u^{(S)}/\partial x^i$.  Then the relevant field equations for a 
Fourier component with wave number $q$ are[8]
\begin{equation} 
\frac{d }{d  t}\Big(a^2\psi\Big)=-4\pi G a^2\Big(\delta 
\rho^{(S)}+3\delta p^{(S)}\Big)\;,
\end{equation} 
and
\begin{equation} 
\delta\dot{\rho}^{(S)}+3H\Big(\delta \rho^{(S)}+\delta p^{(S)}\Big)=-
\Big(\bar{\rho}+\bar{p}\Big)
\,\Big(\psi-a^{-2}q^2 \delta u^{(S)}\Big)\;.
\end{equation} 
Here $\psi$ is a field employed in recent work using synchronous 
gauge[9]
\begin{equation} 
\psi\equiv \frac{d }{d  t}\left(\frac{ 
h_{ii}}{2a^2}\right)\;.
\end{equation}
There is also an Euler equation that will  be needed later in this 
section:
\begin{equation} 
\frac{d }{d  t}\left[a^3\Big(\bar{\rho}+\bar{p}\Big)\delta 
u^{(S)}\right]=-
a^3\delta p^{(S)}\;.
\end{equation}
From equations (45) and (46) together with the relation $
4\pi G \Big(\bar{\rho}+\bar{p}\Big)=-\dot{H}$ it follows that
\begin{equation} 
\dot{A}=-q^2H\delta\dot{u}^{(S)}
\end{equation}
where 
\begin{equation} 
A\equiv a^2H\psi-4\pi G a^2\delta\rho^{(S)} -q^2H\delta u^{(S)}\;.
\end{equation}
Here is the proof: Eq.~(46) can be written
$$ 
\delta\dot{\rho}^{(S)}+3H\Big(\delta \rho^{(S)}+\delta 
p^{(S)}\Big)=\frac{\dot{H}}{4\pi 
G}
\,\Big(\psi-a^{-2}q^2 \delta u^{(S)}\Big)\;,
$$ 
and it follows immediately from Eq.~(45) that
$$
\frac{d  (a^2 H\psi)}{d  t}=-4\pi G a^2 H 
\Big(\delta\rho^{(S)}+3\delta p^{(S)}\Big)+a^2\dot{H}\psi\;.
$$
Eliminating  $\dot{H}\psi$ from these two equations gives
$$
\frac{d  (a^2 H\psi)}{d  t}=-4\pi G a^2 H 
\Big(\delta\rho^{(S)}+3\delta p^{(S)}\Big)
+4\pi G a^2 \left[\delta\dot{\rho}^{(S)}+3H\Big(\delta 
\rho^{(S)}+\delta 
p^{(S)}\Big)\right]
+q^2\dot{H}\delta u^{(S)}\;
$$
or in other words
$$ 
\frac{d }{d  t}\left[a^2H\psi-4\pi G 
a^2\delta\rho^{(S)}\right]=q^2\dot{H}\delta u^{(S)}\;.
$$ 
The quantity in square brackets on the left is 
not invariant under the 
gauge transformations that preserve the condition (43) 
for synchronous 
gauge, so 
instead we work 
with the related gauge-invariant quantity (50), for which Eq.~(49) 
follows immediately.

As long as the velocity potential remains finite in the limit 
$q\rightarrow 0$,
Eq.~(49) yields a conservation law
\begin{equation}
\dot{A}=0~~~~{\rm for}~~q=0\;.
\end{equation} 
This is true for all modes in all cases, including inflation 
with several scalar fields and for the transition from 
radiation to matter 
dominance.  The conservation of $A$ in the limit $q=0$ can also be 
derived by simply perturbing $a(t)$, $\rho(t)$, and the curvature 
constant $K$ in the Friedmann equation, which gives $\delta K=-2A/3$.

By taking suitable linear combinations of solutions, it is always 
possible to 
arrange that for $q=0$ just one of a complete set has $A\neq 0$, while 
all the other solutions have $A=0$.   Examples are given in an 
appendix
to this paper.  Because of the connection of $A$ with the spatial
curvature, it is legitimate to call the solutions with $A=0$ 
isocurvature 
modes. 
When $q$ is small but non-zero the isocurvature solutions 
usually have both $A$ and $\dot{A}$ of order $q^2$, so 
that Eq.~(49) does not keep 
$A$ for these solutions from undergoing large fractional changes.  
This does not vitiate 
the usefulness of the conservation law for initial conditions 
that give a physical perturbation in which all solutions make 
contributions with comparable coefficients.  In this case, the 
contribution of the isocurvature modes to  $A$ may be rapidly 
varying, but at any given time they will be  small as long as $q$  
is sufficiently small.
The physical solution will have a rapid {\em fractional} variation in 
$A$ only if the 
coefficient of the mode with $A\neq 0$ for $q= 0$ is  suppressed, or 
if 
the coefficients of the isocurvature modes are enhanced.  

There is a simple relation between the quantity $A$ introduced in this 
section and the more familiar quantity ${\cal R}$ discussed in 
Sections 
I--III.
Given perturbations $\Psi$, 
$\delta \rho$ and $\delta u$ in Newtonian gauge, we can find 
the perturbations 
$\psi$, $\delta \rho^{(S)}$, and  $\delta u^{(S)}$ in
synchronous gauge from the transformation equations:
\begin{equation}
\psi=-3\dot{\Psi}-
3\frac{d}{dt}\Big(H\epsilon\Big)+\Big(q/a\Big)^2\epsilon\;,~~~~
\delta\rho^{(S)}=\delta\rho-\epsilon\dot{\bar{\rho}}
~~~~\delta u^{(S)}=\delta 
u+\epsilon\;,
\end{equation} 
where 
\begin{equation}
\dot{\epsilon}=\Psi\;.
\end{equation} 
(The possibility of shifting $\epsilon$ by a constant 
term corresponds 
to the possibility of making gauge transformations that 
preserve the 
conditions for synchronous gauge.)  By applying these 
equations to the 
quantity (50), it is elementary to show that
$A$ is related  to the quantity ${\cal R}$ defined in Eq.~(1) by
\begin{equation}
A=-q^2{\cal R}\;.
\end{equation} 
Thus for any finite $q$ the fractional rate of change in 
${\cal R}$ will be the same as 
the fractional rate of change in ${\cal A}$.  In some  treatments 
of multi-field inflation[10] and in discussions of the curvaton 
model[4], it is simply assumed that
the mode with ${\cal R}\neq 0$ and hence $A\neq 0$  is somehow 
suppressed, which is enough to explain why these 
authors find
a significant fractional change in ${\cal R}$.  But why 
more generally does the 
condition $X=0$ play an important role in establishing the 
conservation of ${\cal R}$
for $q\rightarrow 0$ in Newtonian gauge, while there seems to be no 
similar condition needed
for the conservation of $A$ in synchronous gauge?

As a first step toward resolving this apparent paradox, we note 
from Eq.~(54) that that the limit as $q\rightarrow 0$ of the 
perturbed quantities $\delta \rho^{(S)}$ and $\psi$ in synchronous 
gauge in 
the mode for which $A\neq 0$ in this limit  is {\em not} obtained by 
applying a gauge transformation to the  perturbed quantities in the 
corresponding mode in Newtonian gauge for $q=0$, since that would
give $A=0$ for $q=0$.  
We can go further, and show in general that for $q=0$, {\it the 
synchronous gauge solution corresponding to any 
adiabatic solution of the Newtonian gauge field equations (normalized 
to not diverge as $q\rightarrow 0$) has vanishing values not only for 
$A$, but also (up to a choice of a particular synchronous gauge) for 
$\psi$ and the total density fluctuation $\delta \rho^{(S)}$ and 
velocity 
potential $\delta u^{(S)}$.}  

The reasoning here is essentially the reverse
of that used to prove the theorem of Section II.
We will use the space-time component of 
the Einstein field equations in Newtonian gauge
\begin{equation} 
\dot{\Psi}+H\Psi=-4\pi G (\bar{\rho}+\bar{p})\delta 
u=\dot{H}\delta u\;.
\end{equation} 
We work in the limit $q=0$, assuming that the solution is normalized 
so that in Newtonian gauge all fluctuations
remain finite in this limit.  (As we shall see, this assumption is 
less 
innocent than it may seem.)
Then for modes that for $q=0$ are adiabatic in the sense that $X=0$, 
Eqs.~(1) and (2) give 
\begin{equation}
\dot{\Psi}=\frac{d}{dt}\left(H\delta u\right)
\end{equation} 
for $q=0$.  Combining this with Eq.~(55) gives
\begin{equation}
\Psi=-\delta\dot{u}\;.
\end{equation} 
Thus according to Eq.~(53) we can adopt a particular synchronous gauge 
such that the transformation parameter $\epsilon$ in Eq.~(52) is
\begin{equation}
\epsilon=-\delta u\;.
\end{equation} 
Using Eqs.~(56) and (58)  in Eq.~(52) shows immediately that, for 
$q=0$,
\begin{equation}
\psi=0\;.
\end{equation} 
Furthermore, Eq.~(55) together with the Newtonian gauge Euler equation 
supplies a general constraint equivalent to Eq.~(7) for $q=0$:
\begin{equation} 
-4\pi G \delta\rho=3H\dot{H}\delta u
\end{equation} 
Eqs.~(52)  and (60) give the synchronous gauge density fluctuation
\begin{eqnarray}
-4\pi G\delta\rho^{(S)}&=&-4\pi G\left[\delta\rho-
\epsilon\dot{\bar{\rho}}\right]\nonumber\\
&=&3H\dot{H}\delta u+[-4\pi G] \delta u\left[-
3H(\bar{\rho}+\bar{p})\right]\nonumber\\&=&0\;.
\end{eqnarray} 
Finally, the velocity potential in this synchronous gauge is 
\begin{equation}
\delta u^{(S)}= \delta u+\epsilon=0\;.
\end{equation} 

Thus no  synchronous gauge perturbation with non-vanishing 
values of $\psi$ or
$\delta\rho^{(S)}$ or $\delta u^{(S)}$ (apart from those that can be 
eliminated by a transformation to a different synchronous gauge), such 
as modes 1, 2, and 3 of the radiation plus cold dark matter model of 
the 
appendix,  can be the gauge transformation of one of the  $q=0$ 
adiabatic Newtonian gauge solutions.  Rather, the synchronous 
gauge solutions for $q=0$ with $A$ a non-zero constant 
must be the gauge transformations of the terms of 
order $q^2$ in the 
adiabatic Newtonian gauge solution with ${\cal R}\neq 0$,
 re-normalized by dividing by a factor $q^2$.\footnote{This is 
why
it is possible for the quantity $X$ not to vanish in any mode for 
$q=0$
in synchronous gauge, as we find in  the appendix in the case of 
inflation, while
there are two  modes in Newtonian gauge in which $X\rightarrow 0$ for 
 $q\rightarrow 0$,
 despite the fact that $X$ is gauge invariant.  It is not 
that $X$ is 
different in the two gauges, but rather that the limit $q\rightarrow 
0$ means
different things in synchronous and Newtonian gauge.}  With this 
re-normalization of the 
synchronous gauge modes, as in the appendix, the conserved 
quantity $A$ is not necessarily 
of order $q^2$, as would be expected
from Eq.~(54), but can have a finite limit for $q\rightarrow 0$, as we 
will find it does in the appendix.

Now at last we come to the point.  Working in Newtonian gauge, it is 
most natural to assume that, with an over-all normalization factor 
chosen so that ${\cal R}$ is finite and non-zero in the limit $q=0$, 
all density fluctuations and velocity potentials as well as $\Psi$ are 
non-zero in this limit.  Under this assumption, if the contribution of 
non-adiabatic modes is comparable to that of the adiabatic modes, 
${\cal R}$ will undergo significant changes with time.  Transforming 
this sort of solution to synchronous gauge, we have found above
that the density 
fluctuations and the total velocity potential receive contributions of 
order $q^2$ (relative to the Newtonian gauge perturbations) 
from the adiabatic modes but of order unity from the 
non-adiabatic modes, so that $\dot{A}$ is of order $q^2$, while $A$ 
is also of order $q^2$,  and so $A$ does suffer significant 
changes with time.  Or we can re-normalize the synchronous gauge 
fluctuations by an over-all factor of order $1/q^2$, in which case $A$ 
and the density fluctuations and velocity potentials receive 
contributions of order unity for $q=0$ from the adiabatic modes, as in 
the
appendix, while 
the contribution of the non-adiabatic modes to the total velocity
potential if present is enhanced by a 
peculiar looking factor of $1/q^2$, giving both $A$ and $\dot{A}$  
non-zero
limits for $q\rightarrow 0$.

On the other hand, working in synchronous gauge, it is most natural to 
assume that, with an over-all normalization factor chosen so that $A$ 
is finite and non-zero in the limit $q=0$, all density fluctuations 
and 
velocity potentials as well as $\psi$ are finite in this limit.  Under 
this assumption, it makes no difference whether the contribution of 
non-adiabatic modes is comparable to that of the adiabatic modes; even 
if it is,  $A$ will undergo no significant changes with time.  
Transforming this sort of solution to Newtonian  gauge, one finds that 
the density fluctuations and the total velocity potential receive 
contributions of order $1/q^2$ from the adiabatic modes and of order 
unity from the non-adiabatic modes, so ${\cal R}$ is of order $1/q^2$ 
while its rate of change is only of order unity.  Or we can re-
normalize 
the Newtonian gauge fluctuations by an over-all factor of order $q^2$, 
in which case ${\cal R}$ and the density fluctuations and velocity 
potentials receive contributions of order unity for $q=0$, while the 
contribution of any non-adiabatic modes is suppressed by a peculiar 
looking factor of $q^2$, giving ${\cal R}$ a zero rate of change for 
$q=0$.

So which is right?  The issue is not the over-all normalization of the 
total perturbations, but the {\em relative} magnitude of its adiabatic 
and 
non-adiabatic terms in the limit $q\rightarrow 0$.  There is nothing 
about either gauge that makes it a more reliable guide to our 
intuition 
about this than the other.  

It is generally expected that inflation with several scalar fields the 
general
solution does not have $\cal R$ approaching a constant for increasing 
$a(t)$,
in agreement with what would be expected from the behavior for 
$q\rightarrow 0$
suggested by Newtonian gauge but not synchronous gauge.  But there are 
cases of 
multi-field inflation in which $A$ and hence ${\cal R}$ 
do approach  constants as
$a(t)$ increases, as would be expected from the behavior for 
$q\rightarrow 0$
suggested by synchronous gauge but not Newtonian gauge.  One case is a 
potential given by a sum of exponentials[11].  
\begin{equation} 
V=\sum_n g_n\exp(-\lambda_n\varphi_n)
\end{equation} 
Another is a potential of the form
\begin{equation} 
V=F\left(\sum_n\varphi_n^2\right)\;,
\end{equation} 
with $F$ an arbitrary function.
It would be interesting to characterize the general class of 
potentials for
multi-field inflation for which $A$ and  ${\cal R}$ approach constants
as $a(t)$ increases.

\begin{center}
{\bf  APPENDIX: LONG-WAVELENGTH SOLUTIONS IN SYNCHRONOUS GAUGE}
\end{center}
\nopagebreak

In this appendix we will study several examples of calculations for 
zero wave number in
synchronous gauge, to exhibit both solutions with $A\neq 0$ and those 
with $A=0$.
All quantities here will be in synchronous gauge, so we will drop the 
label $(S)$.

As a first example, consider inflation with just a single real scalar 
field $\varphi=\bar{\varphi}(t)+\delta\varphi({\bf x},t)$,  and 
potential $V(\varphi)$.  As is well known, the unperturbed pressure 
and 
energy density are 
\begin{equation} 
\bar{\rho}=\frac{1}{2}\dot{\bar{\varphi}}^2+V(\bar{\varphi})\;,~~~~~~~
\bar{p}=\frac{1}{2}\dot{\bar{\varphi}}^2-V(\bar{\varphi})\;,
\end{equation}
from which we find the equation of motion of the unperturbed scalar 
field
\begin{equation}
\ddot{\bar{\varphi}}+3H\dot{\bar{\varphi}}+V'(\bar{\varphi})=0\;.
\end{equation}
The perturbations to the energy density and pressure are
\begin{equation} 
\delta\rho=\dot{\bar{\varphi}}\delta\dot{\varphi}+V'(\bar{\varphi})
\delta\varphi\;,~~~~~
\delta p=\dot{\bar{\varphi}}\delta\dot{\varphi}-
V'(\bar{\varphi})\delta\varphi\;.
\end{equation} 
Also, the perturbed velocity potential is
\begin{equation} 
\delta u=-\delta\varphi/\dot{\bar{\varphi}}\;.
\end{equation} 
The field equations (45) and (46) for the  Fourier component of the 
perturbations with wave number $q$ here take the form
\begin{equation} 
\frac{d}{d t}\left(a^2\psi\right)=-4\pi 
Ga^2\Big(4\dot{\bar{\varphi}}\,\delta\dot{\varphi}-
2V'(\bar{\varphi})\,\delta\varphi\Big)\;,
\end{equation} 
\begin{equation} 
\delta\ddot{\varphi}+3H\delta\dot{\varphi}+V''(\bar{\varphi})\,\delta
\varphi+a^{-2}q^2\delta\varphi=-\dot{\bar{\varphi}}\,\psi\;,
\end{equation}
where 
\begin{equation} 
H\equiv \frac{\dot{a}}{a}=\sqrt{\frac{8\pi 
G}{3}\left(\frac{\dot{\bar{\varphi}}^2}{2}+V(\bar{\varphi})\right)}\;.
\end{equation}
The Euler equation (48) gives no new information here.
 
There is a gauge mode, with $\varphi=\tau \dot{\bar{\varphi}} $ and 
$\psi=\tau (3\dot{H}-q^2/a^2) $, where
$\tau$ is an arbitrary time-independent function of $q$.  Knowing this 
solution allows us to reduce the degree of equations (67) and (68) 
from 
three to two, in agreement with the number of physical solutions found
in Newtonian gauge in Section III.  We introduce 
time-dependent functions $f$ and $g$ by writing 
\begin{equation} 
\delta\varphi=f\dot{\bar{\varphi}}\;,~~~~~~\psi=(f+g) \Big(3\dot{H}-
q^2/a^2\Big)\;.
\end{equation} 
Equations (67) and (68) then become a second-order set of equations 
for 
the gauge-invariant quantities
$\dot{f}$ and $g$:
\begin{equation} 
\ddot{f}+3H\dot{f}+(\ddot{H}/\dot{H})\dot{f}=-\Big(3\dot{H}-
q^2/a^2\Big)g\;,
\end{equation} 
\begin{equation} 
\Big(3\dot{H}-
q^2/a^2\Big)\dot{g}+\Big(6H\dot{H}+3\ddot{H}\Big)g=\Big(\dot{H}+q^2/a^
2
\Big)\dot{f}\;,
\end{equation} 
in which the gauge mode appears in the possibility of adding a 
constant 
to $f$.  These equations can be solved exactly for $q=0$ and $H(t)$ 
arbitrary.  There are two physical solutions:

\vspace{6pt}

\noindent
{\bf Mode 1:}
\begin{equation} 
g^{(1)}(t)=\frac{1}{3a^3(t)\dot{H}(t)}\int^t_0a(t')dt'\;,~~~~~~\dot{f}
^
{(1)}(t)=\frac{1}{ a^2(t)\dot{H}(t)}\left[1-\frac{H(t)}{a(t)} \int^t_0 
a(t')dt'\right]\;.
\end{equation} 

\vspace{6pt}

\noindent
{\bf Mode 2:}
\begin{equation} 
g^{(2)}(t)=\frac{1}{3a^3(t)\dot{H}(t)}\;,~~~~~~~~~\dot{f}^{(2)}(t)=-
\frac{H(t)}{ a^3(t)\dot{H}(t)}\;.
\end{equation} 

\vspace{6pt}

\noindent
(The lower limit 0 on the integral over $t'$ is arbitrary; changing it 
just amounts to adding some of mode 2 to mode 1.)

Equation (50) gives the values of $A$ for $q=0$ in these two modes as 
the
constants
\begin{equation} 
A_1=1\;,~~~~~~~~A_2=0\;,
\end{equation} 
even though neither of these 
solutions satisfies the  adiabatic condition $X=0$.
A general mixture of modes with coefficients $c_1$ and $c_2$ will have 
$A=c_1$ for $q=0$, provided
$c_2$ does not blow up in this limit.  With this proviso, the 
conservation of $A$ allows the value of $c_1$ that is calculated for a 
given inflaton potential to be used to find the strength of the 
non-isocurvature mode at later times.  However, if the value of $c_2$ 
for 
the physical solution found after horizon crossing  went as $c/q^2$ 
for 
$q\rightarrow 0$, while $c_1$ remained finite, then Eqs.~(49), (66), 
(70), and (74)   would give
$$
\dot{A}\rightarrow c_2\dot{A}_2\rightarrow \frac{-
cH^2(t)}{a^3(t)\dot{H}(t)}~~~{\rm for}~~q\rightarrow 0\;,
$$ 
and there would be no useful conservation law even for $q=0$. 
As discussed in Section IV, this is just what we would expect if we
assumed that, with an over-all normalization factor chosen so
that ${\cal R}$ is finite and non-zero in the limit $q=0$, all
density fluctuations and velocity potentials {\em in Newtonian gauge}
as well as $\Psi$ are
finite and non-zero in this limit.

For another example, we consider a later epoch, when the dominant 
constituents of the  
universe were radiation and cold dark matter.  (For simplicity, we are 
neglecting  the baryon density compared with the density of cold dark 
matter, but supposing that there are  still enough baryons to keep the 
radiation in thermal equilibrium, and we are ignoring the effects of 
free-streaming neutrinos.)  We adopt a particular synchronous gauge in 
which the cold dark matter is at rest.  The field equations then are 
Eqs.~(46) and (48) for the radiation energy density perturbation 
$\delta\rho_R$ and velocity potential $\delta u_R$; Eq.~(46) for the 
cold dark matter density $\rho_D$; and Eq.~(45) for $\psi$, with the 
total energy density and pressure appearing on the right hand side:
\begin{equation}
\delta\dot{\rho}_R+4H\delta\rho_R=-(4/3)\bar{\rho}_R\Big(\psi-a^{-
2}q^2\delta u_R\Big)\;,
\end{equation} 
\begin{equation} 
4\frac{d }{d  t}\left[a^3\bar{\rho}_R\delta u_R\right]=-a^3\delta 
\rho_R\;,
\end{equation} 
\begin{equation}
\delta\dot{\rho}_D+3H\delta\rho_D=-\bar{\rho}_D\psi\;,
\end{equation} 
\begin{equation} 
\frac{d}{dt}\Big(a^2\psi\Big)=-4\pi G 
a^2\left(2\delta\rho_R+\delta\rho_D\right)\;.
\end{equation} 
The unperturbed radiation and dark matter densities go as $a^{-4}$ and 
$a^{-3}$, respectively.  It is convenient here to normalize $a$ so 
that 
$a=1$ when $\bar{\rho}_R=\bar{\rho}_D$, so  that
\begin{equation} 
\bar{\rho}_R=\rho_{EQ}a^{-4}\;,~~~~~~~\bar{\rho}_D=\rho_{EQ}a^{-3}\;,
\end{equation} 
where $\rho_{EQ}$ is constant.

Equations (78)--(81) are a fourth-order system of differential 
equations, so there are four modes, all of which are physical because 
the gauge has been fixed by choosing $\delta u_D=0$.  For $q=0$, they 
take the form

\vspace{6pt}

\noindent
{\bf Mode 1:} 
\begin{eqnarray}
\delta\rho^{(1)}_R&=&\frac{1}{\pi G a^6}\Bigg(16+8a-
2a^2+a^3\Bigg)\;,\nonumber\\
\delta u_R^{(1)}&=&-\frac{a}{4\rho_{EQ}}\sqrt{\frac{3}{8\pi G
\rho_{EQ}}}\int^a \frac{a^4\,da}{\sqrt{1+a}} \delta\rho^{(1)}_R \;,\\
\delta\rho^{(1)}_D&=&\frac{3}{4\pi G a^5}\Bigg(16+8a-
2a^2+a^3\Bigg)\;,\nonumber\\
\psi^{(1)}&=&2\sqrt{\frac{3}{8\pi 
G\rho_{EQ}}}\frac{\sqrt{1+a}}{a^4}\Bigg(32+8a-a^3\Bigg)\nonumber
\end{eqnarray}

\vspace{6pt}

\noindent
{\bf Mode 2:}
\begin{eqnarray}
\delta\rho^{(2)}_R&=&\frac{1}{\pi G a^6}\sqrt{1+a}\;,\nonumber\\
\delta u_R^{(2)}&=&-\frac{a}{4\rho_{EQ}}\sqrt{\frac{3}{8\pi G
\rho_{EQ}}}\int^a \frac{a^4\,da}{\sqrt{1+a}} \delta\rho^{(2)}_R \;,\\
\delta\rho^{(2)}_D&=&\frac{3}{4\pi G a^5}\sqrt{1+a}\;,\nonumber\\
\psi^{(2)}&=&\sqrt{\frac{3}{8\pi G\rho_{EQ}}}\Bigg(4+3a\Bigg)\nonumber
\end{eqnarray}

\vspace{6pt}

\noindent
{\bf Mode 3:}
\begin{eqnarray}
&&\delta\rho^{(3)}_R=\delta\rho^{(3)}_D =\psi^{(3)}=0\;,\nonumber\\
&&\delta u_R^{(3)}\propto a\;. 
\end{eqnarray}

\vspace{6pt}

\noindent
{\bf Mode 4:}
\begin{eqnarray}
\delta\rho^{(4)}_R&=&\frac{1}{\pi G a^6}\Bigg(8+4a-a^2-
8\sqrt{1+a}\Bigg)\;,\nonumber\\
\delta u_R^{(4)}&=&-\frac{a}{4\rho_{EQ}}\sqrt{\frac{3}{8\pi G
\rho_{EQ}}}\int^a \frac{a^4\,da}{\sqrt{1+a}} \delta\rho^{(3)}_R \;,\\
\delta\rho^{(4)}_D&=&\frac{3}{8\pi G a^5}\Bigg(8+4a-
8\sqrt{1+a}\Bigg)\;,\nonumber\\
\psi^{(4)}&=&
\frac{8}{a^4}\sqrt{\frac{3}{8\pi G\rho_{EQ}}}\Bigg((4+a)\sqrt{1+a}-4-
3a\Bigg)\nonumber
\end{eqnarray}

\noindent
The lower bound on the integrals in the formulas for $\delta u_R$ in 
modes 1, 2, and 4 are arbitrary; changing this lower limit in any of 
these integrals just amounts to adding some of mode 3 to that mode.

Note that modes 1, 2, and (trivially) 3 are adiabatic, in the sense 
that 
\begin{equation} 
\frac{\delta\rho_D}{\bar{\rho}_D}=\frac{\delta\rho_R}{\bar{\rho}_R+
\bar{p}_R}\;,
\end{equation} 
(and so $X=0$) while mode 4 is not adiabatic in this sense.
The values of $A$ for the four modes are
\begin{equation} 
A_1=1,~~~A_2=A_3=A_4=0\;.
\end{equation} 
Thus modes 2 and 3 are both adiabatic and isocurvature.  An arbitrary 
mixture of 
modes
will have $A$ constant unless the coefficients of modes 2, 3, or 4 
blow up as 
$1/q^2$ 
limit $q\rightarrow 0$, which will be the case if the fluctuations in 
the 
non-adiabatic modes in Newtonian gauge have non-zero limits for 
$q\rightarrow 
0$.

\begin{center}
{\bf  ACKNOWLEDGMENTS}
\end{center}
\nopagebreak

I am grateful for helpful correspondence with E. Bertschinger, D. 
Lyth,  
S. Mukhanov, and N. Turok. This research was supported in part by the 
Robert A. Welch Foundation and 
NSF Grants PHY-0071512 and PHY-9511632.

\begin{center}
{\bf REFERENCES}
\end{center}

\begin{enumerate}

\item J. M. Bardeen, Phys. Rev. {\bf D22}, 1882 (1980);  
D. H. Lyth, Phys. Rev. {\bf D31}, 1792 (1985).  For reviews, see 
J. Bardeen, in {\em Cosmology and Particle Physics}, eds. Li-zhi Fang 
and A. Zee (Gordon \& Breach, New York, 1988); A. R. Liddle and D. H. 
Lyth, {\em Cosmological Inflation and Large Scale Structure} 
(Cambridge 
University Press, Cambridge, UK, 2000).

\item J. M. Bardeen, Phys. Rev. {\bf D22}, 1882 (1980), Eq.~(5.21). 

\item E. Bertschinger, in {\it Cosmology and Large Scale Structure ---
Proceedings Session LX of the Les Houches Summer School}, ed. R. 
Schaeffer,
J. Silk, M. Spiro, and J. Zinn-Justin (Amsterdam: Elsevier Science, 
1996).

\item S. Mollerach, Phys. Rev. D {\bf 42}, 313 (1990); 
A. D. Linde and V. Mukhanov, Phys. Rev. D {\bf 56}, 
535 (1997); D. H. Lyth
and D. Wands, Phys. Lett. B {\bf 524}, 5 (2002); 
T. Moroi and T. Takahashi, Phys. Lett. B {\bf 522}, 
215 (2001); Phys. Rev.  {\bf D66}, 063501 (2002); 
D. H. Lyth, C. Ungarelli, and D. Wands, astro-ph/0208055; 
K. Dimpopoulos and D. H. Lyth, astro-ph/0209180.

\item J. M. Bardeen, P. J. Steinhardt, and M. S. Turner,
Phys. Rev. {\bf D28}, 679 (1983).  This quantity was 
re-introduced by D. 
Wands, 
K. A. Malik, D. H. Lyth, and A. R. Liddle, Phys. Rev. {\bf D62}, 
043527 
(2000).

\item See, e. g., C. Gordon, D. Wands, B. A. Bassett, and R. Maartens,
Phys. Rev. {\bf D63}, 023506 (2000), Eq.~(14).

\item A. Guth and S-Y. Pi, Phys. Rev. Lett. {\bf 49}, 1110 (1982).

\item These are taken from Eqs.~(15.10.50), (15.10.51), and (15.10.53) 
of S. Weinberg, {\em Gravitation and Cosmology -- Principles and 
Applications of the General Theory of Relativity} (Wiley, New York, 
1972).  It should be noted that the velocity vector ${\bf U}_1$ used 
in 
this reference has components
$U_1^i=a^{-2}\delta u_i=a^{-2}iq_i\delta u^{(S)}$.

\item S. Weinberg, Phys. Rev. {\bf D64}, 123511 
(2001); Phys. Rev. {\bf D64}, 123512 (2001); 
Astrophys. J. {\bf 581}, 810 (2002).

\item See, e.g., V. S. Mukhanov, H. A. Feldman, and R.H. 
Brandenberger, 
Physics Reports {\bf 215}, 203--333 (1992).

\item K. A. Malik and D. Wands, Phys. Rev. {\bf D59}, 123501 (1999).

\end{enumerate}
\end{document}